\documentclass[12pt]{article}
\usepackage{geometry}
\newcommand{\beq}{\begin{equation}}
\newcommand{\eeq}{\end{equation}}
\newcommand{\beqs}{\begin{eqnarray}}
\newcommand{\eeqs}{\end{eqnarray}}
\newcommand{\tr}{\mathrm{tr\,}}
\newcommand{\Q}{{\cal Q}}

\begin{document}

\begin{titlepage}

\hfill ULB-TH/08-32

\vspace{40pt}

\begin{center}
{\Large \bf Supersymmetry and Gravitational Duality}

\end{center}

\vspace{15pt}

\begin{center}
{\large  Riccardo Argurio, Fran\c cois Dehouck and Laurent Houart}\\

\vskip 35pt

Physique Th\'eorique et Math\'ematique and International Solvay Institutes \\
Universit\'e Libre de Bruxelles, C.P. 231, 1050 Bruxelles, Belgium
\end{center}

\vspace{20pt}

\begin{center}
\textbf{Abstract}
\end{center}
We study how the supersymmetry algebra copes with gravitational duality.
As a playground, we consider a charged Taub-NUT solution of
$D=4$, ${\cal N}=2$ supergravity. We find explicitly its Killing spinors,
and the projection they obey provides evidence that the dual magnetic momenta
necessarily have to appear in the supersymmetry algebra.
The existence of such a  modification is further supported
using an approach based
on the Nester form. In the process, we find new expressions for the dual
magnetic momenta, including the NUT charge.
The same expressions are then rederived using gravitational duality.

\end{titlepage}

\section{Introduction}
Supersymmetry has been one of the major ingredients in providing evidence for
dualities in the realm of string theories and M-theory. In particular, there
is a very tight relation between U-duality \cite{Hull:1994ys}, the most
general duality encompassing electric-magnetic duality, S-duality and
T-duality, and the existence of BPS bounds following from the most general
maximally extended supersymmetry algebra. This relation follows from the fact
that states (or supergravity solutions) which preserve some supersymmetries
also saturate a BPS bound which takes the form:
\begin{equation}
M=|Z|, \label{bps}
\end{equation}
where $Z$ is a U-duality invariant combination of all the possible charges
arising in the specific theory one is considering. These charges, which
correspond to possibly extended charged objects, arise in the supersymmetry
algebra as central extensions \cite{de Azcarraga:1989gm,Townsend:1995gp}, and this is the reason why they enter in the BPS bound.

It is however striking that U-duality acts only on the right hand side of the BPS equation (\ref{bps}),
while it leaves the left hand side, $M$, invariant. It is natural to ask
whether there are more general duality transformations that also act on $M$.
Indeed, such a duality exists, at least in four dimensions. It is the gravitational
electric-magnetic duality (see
\cite{Henneaux:2004jw,Deser:2005sz,Bunster:2006rt,Cnockaert:2006gm,
Leigh:2007wf,Bergshoeff:2008vc}
and references therein), which maps the mass $M$ to a magnetic mass $N$,
usually called the NUT charge \cite{ramaswamy,ashtekar}
(see also \cite{Mueller:1985ij,Bossard:2008sw}). It is the purpose of this
paper to study some aspects of this duality in relation to the BPS bound and
the preservation of supersymmetries under it. In particular, in the context of
$D=4$, ${\cal N}=2$ supergravity we discuss how the BPS
equation is generalized in presence of NUT charge to \cite{Kallosh:1994ba}:
\begin{equation}
\sqrt{M^2+N^2}=|Z|, \label{bpsnut}
\end{equation}
and in turn we want to understand how the superalgebra itself takes into
account the possibility of turning on a NUT charge (or more generally a dual
momentum). Some considerations on how the NUT charge transforms
more generally under string dualities have appeared for instance in
\cite{Alvarez:1997yp,Hull:1997kt}.\footnote{We stress that here we
are concerned with the Lorentzian NUT charge. In contrast,
the Euclidean NUT charge, also called the Kaluza-Klein monopole charge,
is extensively discussed in the literature related to string dualities,
where it appears on the same footing as the other $p$-form charges.}

The outline of the paper is as follows. In Section 2, we consider
the Taub-NUT solution of ${\cal N}=2$ supergravity and find
explicitly its Killing spinors, under the condition
(\ref{bpsnut}). In Section 3, we inspect more closely the
projection which defines the Killing spinors. For large radii, it
takes a form which suggests the presence of a new term extending
the supersymmetry algebra, which nevertheless fails to pass the
simple test of hermiticity. In Section 4, we take another route
towards the superalgebra which consists in computing the variation
of the supercharges when expressed in terms of surface integrals.
We recover the same new extension in the r.h.s. of the
superalgebra, but we recognize it now as a ``topological'' term
violating the canonical association of a variation of a surface
charge to the commutator of two such charges. In section 5, we
rederive the expressions for the dual momenta that we obtained in
the previous section, by demanding that they should be defined as
the gravitational duals of the usual ADM momenta. We also show a
way to correctly compute the NUT charge by writing the surface
integrals in such a way that the integrand is free of string-like
singularities. In section 6, we conclude by discussing how one
could reconcile the presence of the dual momenta in the
superalgebra with the theorems that prevent such terms to appear.
In the Appendices we have relegated the conventions and all the
computations that lead to the expression for the Killing spinor.

\section{The BPS Taub-NUT charged solution in ${\cal N}=2$ supergravity}
In this section, we first recall the pure
${\cal N}=2$ supergravity, and display the supersymmetric variation
of the gravitini, which is essentially the Killing spinor equation.
Then, we solve it for an already known solution, namely the black hole
carrying not only mass but also NUT charge (also called ``magnetic mass"), and both electric and
magnetic Maxwell charges as well (see e.g. \cite{AlonsoAlberca:2000cs}).
In the sense of the fall-off conditions used in \cite{Regge:1974zd},
this black hole is asymptotically flat \cite{Bunster:2006rt}
(up to global issues
involving time identifications \cite{Misner:1963fr,Mueller:1985ij}).

The bosonic part of  the ${\cal N}=2$ supergravity Lagrangian is just the
Einstein-Maxwell one:
\begin{equation}
\mathcal{L}= \sqrt{g}\: \biggl [\frac{1}{4} R - \frac{1}{4}
F_{\mu \nu} F^{\mu\nu} \biggl ] \label{action}
\end{equation}
where the signature used is $(-,+,+,+)$ and
$F_{\mu\nu}=2 \: \partial_{[\mu} A_{\nu]}$. We use greek letters
where $\mu,\nu,..=t,r,\theta,\phi$ for curved space indices
and roman letters where $a,b,..=0,1,2,3$ for flat indices.
We essentially follow the conventions of \cite{romans}.

One is usually focusing on supergravity solutions where all fermionic
fields are set to zero. Then, the supersymmetries preserved by such a
solution are simply given by any non-trivial solution to the Killing spinor
equation, which is obtained by setting to zero the supersymmetric variation
of the gravitino spin-$3/2$ field, which is a complex spinor in
${\cal N}=2$ supergravity:
\begin{equation}\label{kilspineq}
\delta{\psi}_{\mu}= \hat{\nabla}_{\mu}\epsilon= \hat{D}_{\mu}
\epsilon + \frac{i}{4} F_{ab} \gamma^{ab}\: \gamma_\mu \: \epsilon
=0
\end{equation}
where $\hat{\nabla}_\mu$ is called the super-covariant
derivative,   and the covariant derivative is
$\hat{D}_{\mu}=\partial_{\mu} +\frac{1}{4}\:
\omega_{\mu}^{\: ab} \gamma_{ab}$.

Here, we take the gamma matrices to be real and such that they satisfy $ \{
\gamma_{a}, \gamma_{b} \}  = 2  \eta_{ab}$. We also have
$\gamma_{ab}=\frac{1}{2}  [\gamma_{a},\gamma_{b}] $. The parity
matrix $\gamma_5$ is real
and antisymmetric $\gamma_5= \gamma_{0123}$.

A special solution to the equations of motion derived
from the action (\ref{action}) is a black hole solution carrying,
besides mass, NUT charge and both electric and magnetic Maxwell charges.
Such a solution is written as:
\begin{eqnarray}
ds^2& = &-\frac{r^2-N^2-2Mr+Q^2+H^2}{r^2+N^2}(dt+2N\cos\theta
d\phi)^2 \nonumber \\
&&+ \frac{r^2+N^2}{r^2-N^2-2Mr+Q^2+H^2} dr^2
+(r^2+N^2)(d\theta^2+\sin^2\theta d\phi^2) , \label{tnmetric}
\end{eqnarray}
\begin{equation}\label{tngauge}
A_{t}=\frac{Qr+NH}{r^2+N^2}, \:\:\:\:\:\:\:
A_{\phi}=\frac{-H(r^2-N^2)+2NQr   }{r^2+N^2}\cos\theta .
\end{equation}
It is easy to see that in the case $N=0$,
we recover the Reissner-Nordstr\"om black hole solution.

Defining
$\lambda=r^2-N^2-2Mr+Q^2+H^2$ and $R^2=r^2+N^2$,
the vielbein read:
\begin{eqnarray}
e^0&=& \frac{\sqrt{\lambda}}{R}(dt+2N\cos\theta d\phi), \:\:
\:\:\:\:\:\: e^1= \frac{R}{\sqrt{\lambda}}dr,
\nonumber \\
e^2&=& Rd\theta,
\:\:\:\:\:\:\:\:\:\:\:\:\:\:\:\:\:\:\:\:\:\:\:\:\:\:\:\:\:\:\:\:\:\:\:\:\:
e^3=R\sin\theta  d\phi. \label{vielbein}
\end{eqnarray}

It is obvious that the Killing spinor equations will have non-trivial
solutions only if the operator acting on the supersymmetry parameter
$\epsilon$ has vanishing eigenvalues, i.e.
its determinant is zero. This will involve a relation among
the constants $M$, $N$, $Q$ and $H$. This relation appears for instance
when computing the integrability conditions of the Killing spinor equations
\cite{AlonsoAlberca:2000cs}. In Appendix A, we provide an alternative
derivation of the same condition.

The BPS condition reads
\beq
M^2+N^2=Q^2+H^2. \label{bpsbound}
\eeq
Note that it is $r$-independent, and that it implies
$\sqrt{\lambda} = r-M$. This is nothing else than
the expression (\ref{bpsnut}), which had already
been derived several years ago in similar contexts
\cite{Kallosh:1994ba,AlonsoAlberca:2000cs}. In order to analyze
in more detail the implications of such a generalized BPS bound,
let us introduce the following expressions:
\beqs
r\pm \gamma_5 N & = & R e^{\pm \beta(r) \gamma_5}, \\
M\pm \gamma_5 N & = & Z e^{\pm \alpha_m \gamma_5}, \\
Q\pm \gamma_5 H & = & Z e^{\pm \alpha_q \gamma_5}
\eeqs
where we have defined $Z^2=M^2+N^2=Q^2+H^2$ and
\beq
\tan \beta= \frac{N}{r}, \qquad \tan \alpha_m = \frac{N}{M}
\qquad \tan \alpha_q = \frac{H}{Q}.
\eeq
Then, the SUSY variations can be rewritten as:
\beqs
\delta \psi_t&=& \partial_t \epsilon + \frac{r-M}{2R^3}Z \gamma_{01}
e^{(\beta-\alpha_m)\gamma_5}\left\{1-ie^{(\beta+\alpha_m-\alpha_q)\gamma_5}
\gamma_0\right\}\epsilon, \\
\delta \psi_r &=& \partial_r \epsilon - \frac{Z}{2R(r-M)} i
e^{(2\beta-\alpha_q)\gamma_5}\gamma_0\epsilon, \\
\delta \psi_\theta &=& \partial_\theta \epsilon -\frac{1}{2} \gamma_{12}
\epsilon + \frac{Z}{2R}\gamma_{12} e^{(\beta-\alpha_m)\gamma_5}
\left\{1-ie^{(\beta+\alpha_m-\alpha_q)\gamma_5}\gamma_0\right\}\epsilon, \\
\delta \psi_\phi &=& \partial_\phi \epsilon
-\frac{1}{2} (\sin\theta \gamma_{13}
+  \cos\theta \gamma_{23})\epsilon + \\
& & + \left[ \frac{Z}{2R} \sin\theta \gamma_{13} + \frac{NZ(r-M)}{R^3}
\cos\theta \gamma_{01}\right] e^{(\beta-\alpha_m)\gamma_5} \nonumber
\left\{1-ie^{(\beta+\alpha_m-\alpha_q)\gamma_5}\gamma_0\right\}\epsilon.
\eeqs

We thus see that it is most natural to look for a Killing spinor
which satisfies the factorization
\begin{eqnarray}
\epsilon(t,r,\theta,\phi)=
e^{\frac{1}{2}\gamma_{12}\theta}\:
e^{\frac{1}{2}\gamma_{23}\phi} \epsilon_0(r), \label{sphkill}
\end{eqnarray}
where $\epsilon_0$ is independent on time
and satisfies the projector equation
\beq
\left\{1-ie^{(\beta+\alpha_m-\alpha_q)\gamma_5}\gamma_0\right\}\epsilon=0.
\label{proj}
\eeq
Note indeed that
\beq
\Pi = \frac{1}{2}\left\{1-ie^{(\beta+\alpha_m-\alpha_q)\gamma_5}
\gamma_0\right\}
\eeq
is a projector, satisfying $\Pi^2=\Pi$. Moreover, since it
verifies $\tr \Pi=2$, it has exactly two zero eigenvalues.

The above result (\ref{proj}) will be essentially enough for the rest
of the discussion on the relation between the Killing spinor and the
supersymmetry algebra. However for the sake of completeness, and in order
to show that a solution indeed exists, we produce below
the complete expression of the Killing spinor.

The only non trivial equation that remains to be solved is $\delta\psi_r=0$.
The final expression for the Killing spinor is (see Appendix B for the
details):
\beq
\epsilon_0(r) =  \left(\frac{r-M}{R}\right)^{\frac{1}{2}}
\left(\begin{array}{c} R[\frac{1}{2}(\beta(r)+\alpha_m-\alpha_q)]
\vec{\epsilon} \\
i R[\frac{1}{2}(\pi -\beta(r)-\alpha_m+\alpha_q)]
\vec{\epsilon} \end{array} \right),
\eeq
where
\beq
\vec{\epsilon} = \left(\begin{array}{c} \epsilon_1 \\ \epsilon_2
\end{array} \right)
\eeq
is a constant two-component complex spinor and
we have defined the rotation matrix
\beq
R[\alpha]= \left(\begin{array}{cc} \cos \alpha & -\sin \alpha \\
\sin  \alpha & \cos  \alpha \end{array} \right).
\eeq

We have thus shown that, provided the extremality
condition (\ref{bpsbound}) is satisfied, the metric has a Killing spinor,
which actually depends on two complex numbers. The metric thus preserves
half of the 8 supersymmetries, as expected from the arguments of
\cite{Tod:1983pm} (see also \cite{Kallosh:1994ba,AlonsoAlberca:2000cs}).

As a last word, we could worry about the issue whether the Killing spinor
is globally defined. Indeed, the metric has a coordinate singularity
along the $z$ axis, also known as the Misner string. One can remove the
singularity along half of the axis by a coordinate transformation.
Essentially, one obtains two completely regular patches
on the upper and lower hemispheres, where the metric is the same
as (\ref{tnmetric}), but with $\cos \theta$ replaced by $\cos\theta \pm 1$.
It amounts to shift the time coordinate $t$ by $\pm 2N\phi$. Since the
Killing spinor is $t$-independent, we can already see that it will be
the same on the two patches. This can be verified by rederiving its expression
as above in the regular metric in each patch. As expected one
finds the same result as above.

\section{The Killing spinor and its asymptotic projection}
In this section we analyze in more detail the solution for the Killing spinor found in the
previous section. In particular, we consider the projection that defines the Killing spinor
and take its limit of large radius, where the metric is asymptotically flat. The projection can be
recast in a form which is similar to the right hand side of the ${\cal N}=2$ supersymmetry
algebra. However, the term containing the NUT charge has the wrong hermiticity condition
and thus does not seem to fit in any of the central (or else) extensions of the most general
${\cal N}=2$ supersymmetry algebra.

The projection defining the four independent real components of the Killing
spinor is given by:
\beq
\left\{1-ie^{(\beta(r)+\alpha_m-\alpha_q)\gamma_5}\gamma_0\right\}\epsilon=0.
\label{projr}
\eeq
We have emphasized that it is $r$-dependent. There are two
observations one can make about this dependence.
Recalling that $\tan \beta(r) = N/r$ and that $\tan \alpha_m =N/M$, we
see that when the NUT charge is absent, both $\beta=0$ and $\alpha_m=0$.
The projector becomes $r$-independent. However, even when $N\neq 0$,
in the limit of large radius, $r\to \infty$, we observe that $\beta \to 0$
and the $r$-dependence disappears. We are thus left with a constant asymptotic
projector which depends on all of the four charges (where it is of course
understood that they satisfy the BPS bound (\ref{bpsbound})).

Let us rewrite the projector in a more readable form. By setting $\beta=0$
and multiplying by $e^{-\alpha_m \gamma_5}$, we obtain:
\beq
\left\{ M-\gamma_5 N -i (Q -\gamma_5 H )\gamma_0\right\}\epsilon=0.
\label{projasymp}
\eeq

We now recall the ${\cal N}=2$ superalgebra, including the scalar central
charges (see e.g. \cite{West:1990tg}).
Using Majorana supercharges $\Q^I$, with $I=1,2$, it is:
\beq
\{\Q^I, \Q^J\} = \gamma^\mu C P_\mu \delta^{IJ}
+ C U^{IJ} + \gamma_5 C V^{IJ},
\eeq
where both $U^{IJ}=-U^{JI}\equiv U \varepsilon^{IJ}$ and
$V^{IJ}=-V^{JI}\equiv V \varepsilon^{IJ}$, and $C$ is the charge
conjugation matrix, which we take here to be $C\equiv \gamma_0$. In our
conventions, Majorana spinors are real. Hence, we can define
a single complex Dirac supercharge:
\beq
\Q = \frac{1}{\sqrt{2}} \left( \Q^1 + i \Q^2 \right).\label{dirac}
\eeq
The only non trivial relation of the superalgebra becomes:
\beq
\{\Q, \Q^\star\} = \gamma^\mu C P_\mu -i (U+ \gamma_5 V) C.
\label{superalg}
\eeq

When there is a multiplet of BPS saturated states,
some combinations of the supercharges have to be represented trivially,
i.e. they have to
vanish. This translates into the statement that the matrix $\{\Q^I, \Q^J\}$,
or equivalently $\{\Q, \Q^\star\}$, is not of maximal rank.
This means that also the right hand side of (\ref{superalg})
must have vanishing eigenvalues.
In the present case, for a massive state at rest, we identify $P_0\equiv M$.
Further, if we set $U\equiv Q$ and $V\equiv H$, we see that
we have preserved supersymmetries if the equation:
\beq
\left\{ M -i (Q -\gamma_5 H )\gamma_0\right\}\epsilon=0
\label{projRN}
\eeq
has solutions (note that we have multiplied the expression
in (\ref{superalg}) by $\gamma^0$ on the left and $C$ on the right).

We recognize the equation (\ref{projasymp}) for $N=0$. So we see that for
a Reissner-Nordstr\"om black hole, the projection on the Killing spinor
in the extremal case maps directly to the right hand side of the
${\cal N}=2$ superalgebra. Actually, we could have
guessed the superalgebra (\ref{superalg}) from the expression for the
projector (\ref{projRN}). It is
thus tempting to do this for the case where $N\neq 0$. From (\ref{projasymp}),
we see that $N$ must belong to a ``charge'' which carries a Lorentz index.
The most straightforward guess is that $N\equiv K_0$ of a vectorial
charge $K_\mu$ which enters the superalgebra as:
\beq
\{\Q, \Q^\star\} \stackrel{?}{=} \gamma^\mu C P_\mu +\gamma_5 \gamma^\mu C K_\mu
-i (U + \gamma_5 V) C.
\label{superalgnut}
\eeq
We see that the NUT charge $N$ seems to belong to an extension of the
superalgebra which is not central in the sense that it is not a Lorentz
scalar. Such extensions have been studied \cite{vanHolten:1982mx},
and the most general
${\cal N}=2$ superalgebra taking them into account has been written
\cite{Ferrara:1997tx,Gauntlett:1999dt}.
It is however straightforward to see that our term with $K_\mu$ is not
part of any extension considered so far. The reason why Eq.(\ref{superalgnut}) is wrong  is extremely simple:
it violates hermiticity. Indeed, we have that
$(\gamma_5 \gamma^\mu C)^\dagger = - \gamma_5 \gamma^\mu C$, while
any term on the right hand side must be hermitian since
$\{\Q, \Q^\star\}^\dagger= \{\Q, \Q^\star\}$.
Before seeking a way to solve this puzzle,
we will see in the following section that $K_\mu$
arises also through a  different argument.

\section{The superalgebra of charges at infinity and the Nester form}
In this section we investigate an alternative approach to understand the existence of the magnetic gravitational charges. We first review the relation between the superalgebra and the
variation of the supercharges when the latter are defined as surface integrals
at spatial infinity \cite{Teitelboim:1977hc,Gibbons:1982fy,Hull:1983ap}. The bosonic charges
appearing in the right hand side of the superalgebra then also appear as
surface integrals at infinity. In this approach, the usual ADM momenta appear
in their covariant formulation, i.e. in terms of the Nester form
\cite{Witten:1981mf,Nester:1982tr}, which is indeed closely related to the
variation of the supercharges. Here we show that, analyzing carefully the
Nester form, the
ADM momenta appear together with the dual, magnetic, ADM momenta.
These charges will appear to be related to  ``topological" terms in the
algebra of the supercharges.  The timelike component of the dual momenta
is nothing
else than the NUT charge discussed previously. Evaluated on the charged NUT
black hole, the right hand side of the superalgebra reduces exactly to the
asymptotic expression contained
in the definition of the Killing spinor, discussed in the previous section.

Let us begin by showing how the Nester form \cite{Nester:1982tr} is related to the variation of the supercharge expressed as a surface integral.
We follow closely \cite{Hull:1983ap}.

Using the Noether method one computes the  generator of supertranslations. It can be written as a volume integral,
which in turn can be expressed as a surface integral:
\beqs
\tilde \Q[\epsilon, \bar \epsilon] &=&\frac{i}{2\pi} \int \varepsilon^{\mu\nu\rho\sigma}
\bar\epsilon  \gamma_5 \gamma_\nu \hat{\nabla}_{\rho} \psi_\sigma d\Sigma_\mu
+c.c. \nonumber \\
&= & -\frac{i}{4\pi} \oint  \varepsilon^{\mu\nu\rho\sigma}
\bar\epsilon  \gamma_5 \gamma_\rho \psi_\sigma d\Sigma_{\mu\nu} + c.c.,
\label{surface}
\eeqs
where $\hat{\nabla}_{\rho}$ is the supercovariant derivative acting on a
spin-3/2 field, $c.c.$ denotes complex conjugate, $\bar\epsilon= \epsilon^\dagger C \equiv \epsilon^\dagger \gamma_0$
and we take the convention $\varepsilon_{0123}=-\varepsilon^{0123}=1$.

The charge $\tilde \Q[\epsilon, \bar \epsilon]$
is bosonic, and it transforms the supergravity fields according
to a supertranslation. When acting for instance on the
bosonic fields, which are real, it generates a variation which is
also real. We recall that  in the present ${\cal N}=2$ case,  the gravitino
$\psi_\mu$ is Dirac and hence complex. In terms of the fermionic Dirac
supercharges defined in (\ref{dirac}) we have:
\beq
\tilde \Q[\epsilon, \bar \epsilon] = i(\bar \epsilon \Q + \bar \Q \epsilon) \label{complex}
\eeq
(note that $(\bar \epsilon \Q)^\star = -\bar \Q \epsilon$).

It follows from the theory of surface charges (see for instance
\cite{Barnich:2001jy,Barnich:2007bf}) that the variation
of the supercharge should define its bracket in the usual way:
\beq
\delta_{\epsilon_1,\bar \epsilon_1}\tilde  \Q[\epsilon_2, \bar \epsilon_2] = i \left[
\tilde \Q[\epsilon_1, \bar \epsilon_1],\tilde \Q[\epsilon_2, \bar \epsilon_2] \right]  \label{algebra0}.
\eeq
In terms of the fermionic supercharges (\ref{dirac}), using  the expression
(\ref{complex}), we would then obtain:
\beq i \left[
\tilde \Q[\epsilon_1, \bar \epsilon_1],\tilde \Q[\epsilon_2, \bar \epsilon_2]
\right]  =
i \bar \epsilon_2 \{ \Q , \Q^\star\} C \epsilon_1 -
i \bar \epsilon_1 \{ \Q , \Q^\star\} C \epsilon_2.\label{total}
\eeq

However we will see that our analysis will force us to consider a possible ``topological extension" namely:
\beq
\delta_{\epsilon_1,\bar \epsilon_1} \tilde \Q[\epsilon_2, \bar \epsilon_2] = i \left[
\tilde \Q[\epsilon_1, \bar \epsilon_1],\tilde \Q[\epsilon_2, \bar \epsilon_2] \right] + T. \label{algebra}
\eeq
The crux of the matter is that
$\delta_{\epsilon_1,\bar \epsilon_1} \tilde \Q[\epsilon_2, \bar \epsilon_2]$
is {\em not} antisymmetric in the exchange of $\epsilon_1$ and $\epsilon_2$,
as we now show.

Using (\ref{surface}) one finds for the bracket term and the ``topological"
term the following expressions
$$
i \left[
\tilde \Q[\epsilon_1, \bar \epsilon_1],\tilde \Q[\epsilon_2, \bar \epsilon_2] \right]
=\frac{1}{2} (\delta_{\epsilon_1,\bar \epsilon_1}\tilde  \Q[\epsilon_2, \bar
\epsilon_2] -\delta_{\epsilon_2,\bar \epsilon_2} \tilde \Q[\epsilon_1, \bar
\epsilon_1]) \qquad \qquad\qquad \qquad \mbox{}
$$
\beq
=-\frac{i}{4\pi} \oint  \varepsilon^{\mu\nu\rho\sigma}
\bar\epsilon_2  \gamma_5 \gamma_\rho \hat{\nabla}_\sigma \epsilon_1   \;
d\Sigma_{\mu\nu} +\frac{i}{4\pi} \oint  \varepsilon^{\mu\nu\rho\sigma}
\hat{\nabla}_\rho \bar\epsilon_1  \gamma_5 \gamma_\sigma  \epsilon_2   \;
d\Sigma_{\mu\nu}- (1 \leftrightarrow 2)\label{variaanti}
\eeq
and
\beqs
T &\equiv & \frac{1}{2} (\delta_{\epsilon_1,\bar \epsilon_1}\tilde  \Q[\epsilon_2, \bar \epsilon_2] +\delta_{\epsilon_2,\bar \epsilon_2}\tilde  \Q[\epsilon_1, \bar \epsilon_1]) \nonumber\\ &=&\frac{i}{4\pi} \oint  \varepsilon^{\mu\nu\rho\sigma}
\hat{\nabla}_\rho( \bar\epsilon_1  \gamma_5 \gamma_\sigma  \epsilon_2 +\bar\epsilon_2  \gamma_5 \gamma_\sigma  \epsilon_1)\; d\Sigma_{\mu\nu} \label{variasym}
\eeqs
Note that obviously (\ref{variaanti}) is identically zero when $\epsilon_1=\epsilon_2$ but $T$ is non-zero.

We now focus on the following expression  which is the ``building
block" of the expressions appearing in (\ref{variaanti})-(\ref{variasym}):
\beq
\hat E^{\mu\nu}\equiv \frac{1}{4\pi}  \varepsilon^{\mu\nu\rho\sigma}
\bar\epsilon  \gamma_5 \gamma_\rho \hat{\nabla}_\sigma \epsilon
\label{nestercom} .
\eeq

This is precisely the expression presented by Nester \cite{Nester:1982tr}
and generalized by Gibbons and Hull \cite{Gibbons:1982fy},
albeit in its complex version\footnote{ In references \cite{Nester:1982tr} and
  \cite{Gibbons:1982fy}, they indeed considered  $\hat E^{\mu\nu}+(\hat
  E^{\mu\nu})^*$.} (recall that $\epsilon$ is Dirac in our
set up). One can see that the (antisymmetric) bracket term (\ref{variaanti})
and the (symmetric) topological term (\ref{variasym}) map respectively to
the real and imaginary parts of the Nester form (\ref{nestercom}).

We are now going to use the expression (\ref{nestercom})
to obtain a linear combination
of purely bosonic surface integrals, which correspond to space-time momenta
and Maxwell charges.
In order to proceed, we
linearize gravity around Minkowski spacetime, in cartesian coordinates.
As we have already seen, we consider space-time endowed with a NUT charge
as asymptotically flat, at least as far as spacelike surface integrals
are concerned \cite{Bunster:2006rt}.

First of all, following \cite{Gibbons:1982fy}, we rewrite the
complex  Nester form as: \beq \hat E^{\mu\nu}= E^{\mu\nu} +
H^{\mu\nu}, \eeq where \beq E^{\mu\nu}= \frac{1}{4\pi}
\varepsilon^{\mu\nu\rho\sigma} \bar\epsilon \gamma_5 \gamma_\rho
\hat{D}_\sigma \epsilon, \qquad H^{\mu\nu} = \frac{i}{16\pi}
\varepsilon^{\mu\nu\rho\sigma} F_{ab} \bar\epsilon \gamma_5
\gamma_\rho \gamma^{ab} \gamma_\sigma \epsilon. \eeq One can
readily check that $H^{\mu\nu}$ is actually real, hence any
surprise will necessarily come from the purely gravitational term
$E^{\mu\nu}$.

In the following, we will express everything in terms of the linearized
spin connection $\omega_{\mu\nu\rho}$.
Hence the covariant derivative on a spinor becomes
(note that we no longer distinguish between flat and
curved indices, since they are the same at first order):
\beq
\hat{D}_\mu \epsilon = \partial_\mu \epsilon
+\frac{1}{4} \omega_{\nu\rho\mu} \gamma^{\nu\rho} \epsilon.
\eeq
We now plug back this expression in $E^{\mu\nu}$. Note that in the
surface integral, the piece proportional to $\partial_\mu \epsilon$
will drop out as explained in detail in \cite{Witten:1981mf,Hull:1983ap}.
The spinors will henceforth be identified with the constant value that
they take asymptotically.\footnote{Indeed, we can actually take the spinors to
be the Killing spinors of flat space in cartesian coordinates.}
Hence we restrict to:
\beq
E^{\mu\nu}=\frac{1}{16\pi}  \varepsilon^{\mu\nu\rho\sigma}
\omega_{\alpha \beta \sigma}
\bar\epsilon  \gamma_5 \gamma_\rho \gamma^{\alpha\beta}\epsilon.
\eeq
Using the relation:
\beq
\gamma_\rho \gamma_{\lambda\tau} =
\eta_{\rho\lambda} \gamma_\tau -\eta_{\rho\tau}\gamma_\lambda
-\varepsilon_{\rho \lambda\tau \xi}\gamma^\xi \gamma_5 ,
\eeq
we thus obtain:
\beqs
E^{\mu\nu}= \frac{1}{8\pi} \bar\epsilon \gamma^\lambda \epsilon
\left( {\omega^{\mu\nu}}_\lambda
+\delta^\mu_\lambda {\omega^{\nu\rho}}_\rho
-\delta^\nu_\lambda {\omega^{\mu\rho}}_\rho \right)
+ \frac{1}{8\pi}  \bar\epsilon \gamma^\lambda\gamma_5 \epsilon\:
\varepsilon^{\mu\nu\rho\sigma} \omega_{\lambda\rho\sigma}.
\eeqs
Note that the first term
above is real while the second is imaginary.

Integrating the above 2-form at spatial infinity, we select the
$E^{0i}$ component, with $i=1,2,3$. We can then reexpress the integral
in terms of purely bosonic surface integrals as:
\beq
\oint E^{0i} d\hat \Sigma_i =
\bar\epsilon \gamma^\lambda P_\lambda \epsilon
+  \bar\epsilon \gamma_5\gamma^\lambda K_\lambda \epsilon ,
\eeq
where we obtain the following expressions for the ADM momenta and the
dual magnetic momenta:
\beqs
P_\lambda & = & \frac{1}{8\pi} \oint ( {\omega^{0i}}_\lambda
+\delta^0_\lambda {\omega^{i\rho}}_\rho
-\delta^i_\lambda {\omega^{0\rho}}_\rho ) d\hat \Sigma_i, \label{admp}\\
K_\lambda & = & \frac{1}{8\pi} \oint
\varepsilon^{ijk} \omega_{\lambda jk}
 d\hat \Sigma_i.\label{admk}
\eeqs
Note that $\varepsilon^{0ijk} = -\varepsilon^{ijk}$.

One can show that the above momenta are such that $P_0=M$ and $K_0
= N$ for the solution (\ref{tnmetric}). We defer to the next
section the discussion of the subtleties of this evaluation along
with the gravitational duality existing between $P_\lambda$ and
$K_\lambda$.

At last, we can also address the second term of the generalized
Nester form, which is treated as in \cite{Gibbons:1982fy}. By writing:
\beq
H^{\mu\nu} = \frac{i}{32\pi} \varepsilon^{\mu\nu\rho\sigma} F_{\lambda\tau}
\bar\epsilon  \gamma_5 (\gamma_\rho \gamma^{\lambda\tau} \gamma_\sigma
-\gamma_\sigma \gamma^{\lambda\tau} \gamma_\rho)\epsilon,
\eeq
and using
\beq
\gamma_\rho \gamma_{\lambda\tau} \gamma_\sigma
-\gamma_\sigma \gamma_{\lambda\tau} \gamma_\rho
= 2\varepsilon_{\rho \lambda\tau \sigma}
\gamma_5 + 2 ( \eta_{\rho\lambda} \eta_{\tau \sigma}
-\eta_{\rho\tau} \eta_{\lambda \sigma} ),
\eeq
we obtain:
\beq
H^{\mu\nu} =\frac{i}{4\pi} F^{\mu\nu} \bar\epsilon\epsilon
+\frac{i}{8\pi} \varepsilon^{\mu\nu\rho\sigma} F_{\rho\sigma}
\bar\epsilon\gamma_5 \epsilon.
\eeq
The surface integral then becomes:
\beq
\oint H^{0i} d\hat \Sigma_i = -i \bar\epsilon U \epsilon
-i \bar\epsilon\gamma_5 V \epsilon,
\eeq
with the central charges defined by:
\beqs
U & = & -\frac{1}{4\pi} \oint F^{0i} d\hat \Sigma_i,\\
V & = & \frac{1}{8\pi} \oint \varepsilon^{ijk} F_{jk} d\hat
\Sigma_i.\label{magncharge} \eeqs It can be checked that $U=Q$ and
$V=H$ on our solution.

Summing up all the terms, we have:
\beq
\oint \hat E^{0i} d\hat \Sigma_i
=\bar\epsilon \gamma^\lambda P_\lambda \epsilon
+  \bar\epsilon\gamma_5 \gamma^\lambda K_\lambda \epsilon
-i \bar\epsilon U \epsilon
-i \bar\epsilon\gamma_5 V \epsilon \label{finalex}.
\eeq
It is clear that the above expression cannot be equated to
$\bar \epsilon \{ \Q, \Q^\star\} C \epsilon$, which would then result
in the ``wrong'' superalgebra (\ref{superalgnut}).
But now we see that the obstruction to do so is precisely the presence
of the topological term $T$ in (\ref{algebra}).

Using the definitions of $T$ (\ref{variasym}) and of  the complex Nester form
(\ref{nestercom}) we see that
\beq
T(\epsilon, \bar{\epsilon}) =-i \oint (\hat E -\hat E^*)
\label{relnecharge}.
\eeq
Using then the result (\ref{finalex}) we finally indeed find:
\beq
T(\epsilon, \bar{\epsilon})= -2 i \bar\epsilon\gamma_5 \gamma^\lambda K_\lambda \epsilon.
\label{topo}
\eeq

To sum up, we see that a refined analysis of the Nester form in its complex version permits to
 recover precisely the additional term which was guessed from the
asymptotic projection acting on the Killing spinor.
In this context, we see that this additional  term  is actually violating the
 relation (\ref{algebra0}) and corresponds to a ``topological" term leading to
 the bosonic algebra (\ref{algebra}). It would be interesting, but beyond
 the scope of this note,
to understand better under the lines of \cite{Barnich:2007bf}, the
 appearance of such  topological terms.

\section{The dual magnetic momenta and a generalization of the ADM formula}
In this section we derive the expressions for the dual magnetic
ADM momenta, containing as the timelike component the NUT charge.
The derivation is based on a straightforward application of the
usual ADM argument (see e.g. \cite{Arnowitt:1962hi}) to the dual
Riemann tensor, in its linearized form. We stress that we express
all the quantities in terms of the linearized spin connection, so
that the Bianchi identities are not automatically satisfied.
Eventually we reformulate the classical treatment of
\cite{Bunster:2006rt} (where the magnetic charge would be obtained
from contributions of the metric and the Misner string) by using
the gauge-variance of the spin connection. We rewrite the
integrals in terms of the vielbein in a fixed gauge so as to
express the surface charges in terms of a regular spin connection,
i.e. without string-like singularities.

In electromagnetism, when magnetic charges are considered, one has
to add a magnetic current to the Bianchi identity. The conserved
magnetic charge is calculated using (\ref{magncharge}). Obviously
this charge would be trivially zero if the field strength verified
$F=dA$ but one has to write $F=dA+C$ where $C$ represents the
contribution from the Dirac string of the monopole to obtain the
magnetic charge. As explained in \cite{Bunster:2006rt}, the
situation looks quite similar in gravity. The Bianchi identities
can be rewritten in terms of the dual Riemann tensor defined by
\beq \tilde R_{\mu\nu\rho\sigma} =  \frac{1}{2}
\varepsilon_{\mu\nu\alpha\beta}
{R^{\alpha\beta}}_{\rho\sigma}\label{gravd} \eeq  in the form \beq
\tilde G^{\mu\nu}=8\pi\Theta^{\mu \nu} \label{gravdual} \eeq where
$\Theta^{\mu \nu}$ is the conserved magnetic stress-energy tensor.

To recover the expression for the NUT charge, let us first begin
by briefly recalling how to quickly obtain the expressions for the
ADM momenta. In this section, all the curvature tensors are to be
considered as the linearized ones. By considering the higher order
terms as belonging to the stress-energy tensor, one arrives at the
definition: \beq P_\mu = \frac{1}{8\pi} \int G_{0\mu}d^3V ,
\label{adm} \eeq which involves a volume integral.

The linearized Riemann tensor is written in terms of the spin
connection: \beq R_{\mu\nu\rho\sigma} = \partial_\rho
\omega_{\mu\nu\sigma} -\partial_\sigma
\omega_{\mu\nu\rho}.\label{linRiem} \eeq We thus have the
following expressions: \beqs \label{g00}
G_{00} &=& \frac{1}{2} R_{ijij} = \partial_i \omega_{ijj} \\
\label{goi}
G_{0i} & = & R_{0jij}  = \partial_i \omega_{0jj} - \partial_j \omega_{0ji}.
\eeqs
Using the definition (\ref{adm}) and rewriting the $P_\mu$ in terms of surface
integrals using (\ref{g00}) and (\ref{goi}) one recovers the usual expression
(\ref{admp}) for the ADM momenta.

Looking at (\ref{gravdual}) and (\ref{adm}) it is natural to
consider that the definition of the conserved magnetic charge is:
\beq K_\mu = \frac{1}{8\pi} \int \tilde G_{0\mu}d^3V .
\label{dualadm} \eeq The dual Ricci tensor is: \beq \tilde
R_{\mu\rho} = \eta^{\nu\sigma} \tilde R_{\mu\nu\rho\sigma} =
\frac{1}{2} \eta^{\nu\sigma} \varepsilon_{\mu\nu\alpha\beta}
{R^{\alpha\beta}}_{\rho\sigma}. \eeq The dual Ricci scalar and
dual Einstein tensor are defined just as $\tilde
R=\eta^{\mu\rho}\tilde R_{\mu\rho}$ and $\tilde G_{\mu\rho} =
\tilde R_{\mu\rho} -\frac{1}{2} \eta_{\mu\rho}\tilde R$. We thus
have the following expressions: \beqs \label{tg00}
\tilde G_{00} &=& -\frac{1}{2} \varepsilon_{ijk} R_{0ijk}=\varepsilon_{ijk}\partial_i \omega_{0jk}, \\
\label{tg0j}
\tilde G_{0i} &=& \frac{1}{2} \varepsilon_{jkl} R_{klij}=
\frac{1}{2} \varepsilon_{jkl}(\partial_i \omega_{klj} - \partial_j \omega_{kli})
=\varepsilon_{jkl} \partial_l \omega_{ijk}.
\eeqs
In the last equality of (\ref{tg0j})
 we have used the identity $\partial_{[i}\omega_{jkl]}=0$.
Note also that $\tilde G_{0i}\neq \tilde G_{i0}$ for an arbitrary
(i.e. off-shell) spin connection.
 Using now the
 definition (\ref{dualadm}) for the dual momenta and using (\ref{tg00})
 and (\ref{tg0j}) we recover exactly (\ref{admk}).
We have thus gained confidence that the expressions that we obtained
through the complex Nester form are indeed what one would expect
from a canonical definition of the dual ADM momenta.

We now want to express our formulas in function of the vielbein.
Here this should be done  carefully. Indeed, there is a big
difference with respect to the case of electromagnetism where the
quantity entering the surface integral is $F_{\mu\nu}$, a gauge
invariant quantity. Since the field strength for a magnetic
monopole is related by duality to the field strength of an
electric charge it will never contain string contributions, this
being obviously true in whatever gauge. For this reason, the
calculation of $F_{\mu\nu}$ can also quickly be done by taking
derivatives of the gauge potential away from the singularities. As
we have just seen, in General Relativity, the conserved magnetic
charge is expressed in terms of a spin connection which is a gauge
variant object. To treat it correctly, one should then express the
spin connection as a function of the metric and the string
contributions, as shown in \cite{Bunster:2006rt}. The calculation
of the NUT charge will then involve contributions of the Misner
string at infinity. However, if we look at (\ref{gravd}) and
(\ref{linRiem}) we see that: \beq \tilde
\omega_{\mu\nu\sigma}=\frac{1}{2}
\varepsilon_{\mu\nu\alpha\beta}\: {\omega^{\alpha
\beta}}_{\sigma}.\label{spindual}\eeq Because the Schwarzschild
metric has no singularities at infinity, this means that there
exists a fixed gauge where the dual metric (the NUT metric) has a
regular spin connection. From now on, all expressions will be
written by supposing that we are in the gauge where the spin
connection is regular at infinity, and derivatives are taken away
from the singularity.

The linearization of the vielbein is: \beq
e^\mu=dx^\mu+\frac{1}{2}
\eta^{\mu\nu}(h_{\nu\rho}+v_{\nu\rho})dx^\rho, \eeq where
$h_{\nu\rho}=h_{\rho\nu}$ and $v_{\nu\rho}=-v_{\rho\nu}$. We
recall that the linearized vielbein has 16 independent components,
while the linearized metric has only 10, precisely the
$h_{\nu\rho}$ above. The extra 6 components $v_{\nu\rho}$ are of
course related to the local Lorentz invariance introduced by the
tetrad formalism. The spin connection then reads: \beq
\label{gaugeconn} \omega_{\mu\nu}=\omega_{\mu\nu\rho}e^\rho,
\qquad \omega_{\mu\nu\rho}= \frac{1}{2}(\partial_\nu h_{\mu\rho}
-\partial_\mu h_{\nu\rho} +\partial_\rho v_{\nu\mu}). \eeq

We now rewrite the charges (\ref{admp}) and (\ref{admk}) using the above
expression. This leads to the following generalized ADM and dual ADM
formulae:
\beqs
P_0 & = & \frac{1}{16\pi} \oint ( \partial_i h_{li}-  \partial_l h_{ii}+ \partial_i v_{il}) d\hat \Sigma_l, \label{p0v}\\
P_k & = & \frac{1}{16\pi} \oint ( \partial_0 h_{lk}- \partial_l h_{0k}
+ \delta^k_l \partial_{i} h_{0i} - \delta^k_l \partial_{0} h_{ii} +\partial_{k} v_{0 l}+ \delta^k_l \partial_{i} v_{i 0} ) d\hat \Sigma_l, \label{pkv}\\
K_0 & = & \frac{1}{16\pi} \oint  \varepsilon^{lij}  ( \partial_i h_{0j}+ \partial_j v_{i0} )  d\hat \Sigma_l , \label{k0v}\\
K_k & = & \frac{1}{16\pi} \oint  \varepsilon^{lij}  ( \partial_i
h_{kj}+ \partial_j v_{ik} )  d\hat \Sigma_l , \label{kkv} \eeqs
where the gauge is fixed such that the spin connection is regular
at infinity and string contributions disappear from the surface
integrals. In order to evaluate the NUT charge, let us first
display the metric as a first order correction to the flat metric
in cartesian coordinates.\footnote{One might be worried by higher order corrections which could  cease being subleading near the string-like singularity. Such terms are quadratic or higher in the charge $N$ (and possibly $M$, $Q$ and $H$), and hence must cancel among themselves in the field equations, since the sources are linear in the charges. As a result, they do not contribute to the surface integrals, as can be checked explicitly in the case below if we were to retain also the higher order terms in the metric.}
{}For simplicity, and since the different
perturbations are independent, we set the mass parameter $M$ to
zero: \beq ds^2 = -dt^2 + dx^2 +dy^2 + dz^2 +4N \frac{z}{r}
\frac{ydx-xdy}{x^2+y^2}dt +{\cal O}(N^2). \eeq Following
(\ref{vielbein}), a natural choice for the vielbein is the
triangular one: \beq e^t = dt -2N \frac{zy}{r(x^2+y^2)}dx + 2N
\frac{zx}{r(x^2+y^2)}dy, \eeq \beq e^x= dx, \qquad e^y=dy, \qquad
e^z = dz. \eeq In terms of the tensors $h_{\mu\nu}$ and
$v_{\mu\nu}$, we have: \beq h_{tx}=v_{tx}= 2N
\frac{zy}{r(x^2+y^2)}, \quad h_{ty}=v_{ty}= -2N
\frac{zx}{r(x^2+y^2)}. \eeq Note that when $M\neq 0$, the tensor
$v_{\mu\nu}$ only depends on $N$ at the linear level (i.e., the
$M$-dependent perturbation of the vielbein is symmetric). Hence
its presence is not going to affect the computation of the ADM
mass $P_0$. On the other hand, in the expression giving $P_i$ it
can be checked that its presence makes the integrand vanishing.
The charges $K_i$ also straightforwardly vanish. We are left with
computing $K_0$.

By choosing a particular vielbein, we fixed the gauge. Evaluating
the linearized spin connection, using for instance
(\ref{gaugeconn}) one finds: \beq \omega_{0ij}=-\omega_{ij0}=
\varepsilon_{ijk}\frac{Nx^k}{r^3}. \eeq  An additional check of
the validity of this particular gauge comes from the fact that one
recovers the same spin connection if calculated using the dual of
the linearized spin connection for the Schwarzschild metric by
means of (\ref{spindual}). In other words, we have used the local
Lorentz gauge freedom of the vielbein to obtain a regular spin
connection. In some sense, the Misner string has been gauged away,
and the expressions for the surface charges given above become
completely reliable.
At last, using the above expression it is straightforward to get:
\beq
K_0 =  \frac{1}{8\pi} \oint
\varepsilon^{ijk} \omega_{0jk}
 d\hat \Sigma_i = \frac{N}{4\pi} \oint \frac{x^i}{r^3}d\hat \Sigma_i = N.
\eeq

We have thus shown that the surface charges computing $M$ and $N$
for the (charged) Taub-NUT metric are indeed $P_0$ and $K_0$
respectively, and that they can be both obtained from the Nester
form and independently from an ADM-like argument involving the
dual Riemann tensor.

\section{Discussion}
In this final section we discuss the results we have derived in the previous
sections. Taub-NUT spaces are notoriously problematic for the time
identifications that they imply \cite{Mueller:1985ij},
and for the presence of the Misner strings \cite{Misner:1963fr},
which are gauge-variant singularities. It has been suggested that these
pathologies are enough to conclude that such spacetimes are not globally
supersymmetric \cite{Ortin:2006xg},
even though they have locally (and globally as well)
Killing spinors. However from the point of view of the surface integrals that
define both the bosonic
and the fermionic charges of the superalgebra, the spacetime with NUT charge
is asymptotically flat according to the simplest definition
\cite{Bunster:2006rt}. If we were to assume
that the presence of Killing spinors implies that the spacetime is
supersymmetric, we would be faced with the challenge of including the NUT
charge in 
the superalgebra. The (asymptotic) projection acting on the Killing spinor
must be the same as the projection acting on the supercharges which are
represented trivially on a BPS multiplet. However as we have shown the NUT
charge enters in a term which cannot be part of the superalgebra because of
its wrong hermiticity. Below, we suggest a tentative path to
trivialize this problem.

A logical possibility is to write the corrected variation of the supercharge
(\ref{algebra}) in a different form, by introducing a new supercharge
$\tilde \Q'$:
\beq
\delta_{\epsilon_1,\bar \epsilon_1} \tilde \Q[\epsilon_2, \bar \epsilon_2] = i \left[
\tilde \Q'[\epsilon_1, \bar \epsilon_1],\tilde \Q[\epsilon_2, \bar \epsilon_2] \right]. \label{algebranew}
\eeq
The above expression is not antisymmetric under the exchange of
$\epsilon_1$ and $\epsilon_2$, which is another way of encoding the presence
of the (symmetric) topological term.
In terms of the fermionic supercharges $\Q$ and $\Q'$, (\ref{algebranew})
reads:
\beq
\delta_{\epsilon_1,\bar \epsilon_1} \tilde \Q[\epsilon_2, \bar \epsilon_2] =
i \bar \epsilon_2 \{ \Q , {\Q'}^\star\} C \epsilon_1 -
i \bar \epsilon_1 \{ \Q' , \Q^\star\} C \epsilon_2,\label{totalnew}
\eeq
where we have supposed that $\{\Q,\Q'\}=0$.
Then, equating the above to the expression obtained through the
Nester form, we get:
\beq
\{\Q, {\Q'}^\star\} = \gamma^\mu C P_\mu +\gamma_5 \gamma^\mu C K_\mu
-i (U + \gamma_5 V) C.
\label{superalgnutnew}
\eeq
Now the l.h.s. is no longer hermitian, so there are no obstructions
to having the antihermitian term containing $K_\mu$ in the r.h.s.
The question is of course what is $\Q'$. It must be related to $\Q$
otherwise we would be doubling the number of supercharges.\footnote{Notice the
similarity with pseudo-supersymmetry (see e.g. 
\cite{Mostafazadeh:2002sk,Townsend:2007aw}), where one is also
dealing with non-hermitian relatives of superalgebras.}
We now show
that it is related to $\Q$ through an ``axial" phase shift.

Let us rewrite for definiteness the relation (\ref{superalgnutnew}) on
our particular static massive, charged states with NUT charge:
\beq
\{\Q ,  {\Q'}^\star \} = M+\gamma_5 N -i (Q +\gamma_5 H )\gamma_0.
\label{proposal}
\eeq
Using the angles defined in Section 2, it can be rewritten
as:
\beq
\{\Q , {\Q'}^\star\} = \sqrt{M^2+N^2} e^{\alpha_m \gamma_5}
-i Z e^{\alpha_q \gamma_5} \gamma_0.
\eeq
If the charge
$\Q'$ is related to  $\Q$ by a simple
phase rotation:
\beq
{\Q'}^\star = \Q^\star e^{\alpha_m \gamma_5},\label{phase}
\eeq
then eq.~(\ref{proposal}) takes a more standard, hermitian form:
\beq
 \{\Q , \Q^\star \} = M' -i (Q' +\gamma_5 H' )\gamma_0,
\label{new}
\eeq
with
\beq
M'=
\sqrt{M^2+N^2},\qquad  Q'= \frac{QM-HN}{\sqrt{M^2+N^2} }, \qquad
H'=  \frac{HM+QN}{\sqrt{M^2+N^2} }.
\eeq
Hence, through a non-linear redefinition of the charges, we obtain the
relation (\ref{new}) that in the new variables defines an hermitian
superalgebra. Actually, the new variable $M'$ is precisely the result of a
gravitational duality rotation that eliminates the NUT charge, namely:
\beq
\left( \begin{array}{cc} \cos\alpha_m & \sin \alpha_m \\
-\sin \alpha_m &  \cos\alpha_m  \end{array}\right)
\left( \begin{array}{c} M \\ N  \end{array}\right) = 
\left( \begin{array}{c} M' \\ 0  \end{array}\right).
\eeq
Note that also $Q'$ and $H'$ are obtained from $Q$ and $H$ through an
electromagnetic duality rotation of the same angle.

The phase rotation (\ref{phase}) depends on dynamical quantities,
such as $N$ and $M$. The latter however commute with
the supercharges for consistency of the superalgebra, hence for instance
we are assured that $\{\Q , \Q' \} = 0$. Moreover, one could wonder what
modified supersymmetry variation is induced by $\Q'$. This clearly deserves
to be investigated, though for consistency we anticipate that we should not
find any modification in the transformation
laws of the elementary fields.

In a more
general case where both ordinary and NUT momenta $P_i$ and $K_i$ are non zero
the situation is a bit subtler.
Indeed, focusing only on the ``gravitational'' part, we would have:
\beq
\{\Q , {\Q'}^\star\} = P_0 +\gamma_5 K_0 +
(P_i + \gamma_5 K_i ) \gamma^i\gamma_0.
\eeq
After a rotation similar to (\ref{phase}) we would get:
\beq
 \{\Q , \Q^\star \} = \sqrt{P_0^2+K_0^2} + \frac{1}{\sqrt{P_0^2+K_0^2}}
\left[ P_i P_0 + K_i K_0 + \gamma_5 ( K_i P_0 - P_i K_0) \right]
\gamma^i\gamma_0.\label{offending}
\eeq
We thus still have an offending anti-hermitian term, which is however 
proportional to $K_i P_0 - P_i K_0$ and is thus not present when $K_\mu$
is parallel to $P_\mu$. Now, under a general gravitational duality rotation 
\cite{Bunster:2006rt}
we have that:
\beq
\left( \begin{array}{cc} \cos\alpha & \sin \alpha \\
-\sin \alpha &  \cos\alpha  \end{array}\right)
\left( \begin{array}{c} P_\mu \\ K_\mu  \end{array}\right) = 
\left( \begin{array}{c} P_\mu' \\ K'_\mu  \end{array}\right),
\label{gravrot}
\eeq
and a NUT 4-momentum $K_\mu$ can be completely eliminated only if it is
parallel to $P_\mu$. We thus seem to be able to make sense out of a
superalgebra in the presence of NUT charges only when the latter can be
eliminated by a gravitational duality rotation.

When this is not possible, we do not seem to be able to define a superalgebra.
Note that we are not aware of solutions with non-aligned $K_\mu$ and
$P_\mu$ charges. Actually, it can be shown on simple examples that the
r.h.s. of (\ref{superalgnutnew}) does not have vanishing eigenvalues when
$K_\mu$ and $P_\mu$ are non parallel.

In the case $K_\mu = \lambda P_\mu$, we have $\lambda=N/M =
\tan \alpha_m $ and performing the rotation (\ref{gravrot}) with 
$\alpha=\alpha_m$, the relation (\ref{offending}) becomes the usual
superalgebra:
\beq
 \{\Q , \Q^\star \} =\gamma^\mu C P_\mu' .
\eeq

Note that $K_\mu$ is always parallel to $P_\mu$
if the spatial components $K_i$ and $P_i$ are
obtained by boosting a static object with $K_0$ and $P_0$ charges.
We show in \cite{adh2} that boosting a pure Taub-NUT solution,
one indeed obtains solutions with $K_i\neq 0$, and that in the infinite
boost limit, one recovers the magnetic dual of the usual pp-wave,
which is moreover half-BPS. This latter fact lends support to
the presence of the dual magnetic momenta even in the ${\cal N}=1$
superalgebra, along the same lines as above.

We could thus sum up in the following way the answer to the question that
motivated this work, namely how does the NUT charge enter in the supersymmetry
algebra. When $K_\mu$ is parallel to $P_\mu$, which seems to be the only
situation where we have Killing spinors, by a gravitational duality rotation
(\ref{gravrot}) we can eliminate $K_\mu$. The superalgebra then incorporates
the NUT charges through the (duality invariant) combination $P_\mu'$. 
Alternatively, we can define a generalization of the superalgebra 
(\ref{superalgnutnew}) where the NUT charges appear on the r.h.s. but where 
we have to define a new supercharge through the axial phase rotation
(\ref{phase}). It is this latter generalized superalgebra that can be directly
related to the complex Nester form. Nevertheless, 
both alternatives give the same BPS bound and projection
on the supercharges, and are hence compatible with the projection on the
Killing spinor.  In conclusion, this is evidence that backgrounds which are
obtained 
through gravitational duality rotations from ordinary BPS solutions, such as
Reissner-Nordstr\"om black holes, are indeed supersymmetric.

\subsection*{Acknowledgments}

We are greatly indebted to Glenn Barnich and C\'edric Troessart  for many interesting and
fruitful  discussions.
We would also  like to thank Marc Henneaux, Chris Hull, Axel Kleinschmidt and Tomas Ort\'{\i}n for interesting conversations.

This work was supported in part by IISN-Belgium (conventions
4.4511.06, 4.4505.86 and 4.4514.08) by the European Commission FP6 RTN
programme MRTN-CT-2004-005104, and by
the Belgian Federal Science Policy Office through the
Interuniversity Attraction Pole P5/27. R.A. and L.H. are
Research Associates of the Fonds
de la Recherche Scientifique--F.N.R.S. (Belgium).

\appendix

\section{Computation of the variation of the gravitino}
In this Appendix, we compute the variation of the gravitino,
which is a complex Dirac spinor in
${\cal N}=2$ supergravity:
\begin{equation}\label{kilspineq2}
\delta{\psi}_{\mu}= \hat{\nabla}_{\mu}\epsilon= \hat{D}_{\mu}
\epsilon + \frac{i}{4} F_{ab} \gamma^{ab}\: \gamma_\mu \: \epsilon
=0
\end{equation}
where we recall that $\hat{\nabla}_\mu$ is the super-covariant
derivative  and
$\hat{D}_{\mu}=\partial_{\mu} +\frac{1}{4}\:
\omega_{\mu}^{\: ab} \gamma_{ab}$.

We take the gamma matrices to be real and such that they satisfy $ \{
\gamma_{a}, \gamma_{b} \}  = 2  \eta_{ab}$. We also denote
$\gamma_{ab}=\frac{1}{2}  [\gamma_{a},\gamma_{b}] $.
$\gamma_5= \gamma_{0123}$ is real and antisymmetric.
For definiteness, we list below a choice of real gamma matrices:
\begin{eqnarray}
\gamma_0 &=& \left (
\begin{array}{cccc}
0 & 0 & 0  & -1 \\
0 & 0 & 1  &  0 \\
0 &-1 & 0  &  0 \\
1 & 0 & 0  &  0
\end{array}
\right ) \:\:\: \gamma_1= \left (
\begin{array}{cccc}
1 & 0 & 0  & 0 \\
0 & -1 & 0  &  0 \\
0 & 0 & 1  &  0 \\
0 & 0 & 0  &  -1
\end{array}
\right ) \nonumber \\
 \gamma_2 &=& \left (
\begin{array}{cccc}
0 & 0 & 0  & 1 \\
0 & 0 & -1  &  0 \\
0 & -1 & 0  &  0 \\
1 & 0 & 0  &  0
\end{array}
\right ) \:\:\: \gamma_3= \left (
\begin{array}{cccc}
0 & -1 & 0  & 0 \\
-1 & 0 & 0  &  0 \\
0 & 0 & 0  &  -1 \\
0 & 0 & -1  &  0
\end{array}
\right )
\end{eqnarray}
We use conventions where $C=\gamma_0$, $\bar \epsilon=\epsilon^\dagger C$
and $\varepsilon_{0123}=-\varepsilon^{0123}=1$.

Using the definitions $\lambda=r^2-N^2-2Mr+Q^2+H^2$ and $R^2=r^2+N^2$,
the charged Taub-NUT solution that we study is:
\begin{equation}
ds^2 = -\frac{\lambda}{R^2}(dt+2N\cos\theta
d\phi)^2 \nonumber
+ \frac{R^2}{\lambda} dr^2
+R^2(d\theta^2+\sin^2\theta d\phi^2) , \label{tnmetric2}
\end{equation}
\begin{equation}\label{tngauge2}
A_{t}=\frac{Qr+NH}{R^2}, \:\:\:\:\:\:\:
A_{\phi}=\frac{-H(r^2-N^2)+2NQr   }{R^2}\cos\theta .
\end{equation}

We choose the vielbein to be:
\begin{eqnarray*}
e^0&=& \frac{\sqrt{\lambda}}{R}(dt+2N\cos\theta d\phi), \:\:
\:\:\:\:\:\: e^1= \frac{R}{\sqrt{\lambda}}dr,
\nonumber \\
e^2&=& Rd\theta,
\:\:\:\:\:\:\:\:\:\:\:\:\:\:\:\:\:\:\:\:\:\:\:\:\:\:\:\:\:\:\:\:\:\:\:\:\:
e^3=R\sin\theta  d\phi.
\end{eqnarray*}
We also list below the non-trivial
components of the spin connection:
\begin{eqnarray*}
\omega_{t}^{\:\:01}&=&
\frac{\lambda'}{2R^2}-\frac{\lambda}{R^3}R'\:\:\:\:\:\:\:\:\:\:\:\:\:\:\:\:\:\:\:\:\:\:\:\:\:
\omega_{\theta}^{\:\:12}=-\frac{\sqrt{\lambda}}{R}R' \\
 \omega_{\phi}^{\:\:13}&=& -\frac{\sqrt{\lambda}}{R}R'\sin\theta
\:\:\:\:\:\:\:\:\:\:\:\:\:\:\:\:\:\:\:\:\:\:\:\:\:
 \omega_{\phi}^{\:\:23}=- \cos\theta (1+\frac{2\lambda N^2}{R^4})
 \\
 \omega_{\phi}^{\:\:02}&=&
 -\frac{\sqrt{\lambda}}{R^2}N\sin\theta \:\:\:\:\:\:\:\:\:\:\:\:\:\:\:\:\:\:\:\:\:\:\:\:\:
  \omega_{\theta}^{\:\:03}=\frac{\sqrt{\lambda}}{R^2}N \\
   \omega_{t}^{\:\:23}&=&-\frac{\lambda}{R^4}N \:\:\:\:\:\:\:\:\:\:\:\:\:\:\:\:\:\:\:\:\:\:\:\:\:\:\:\:\:\:\:\:\:\:\:
\omega_{\phi}^{\:\:01}=2N\cos\theta
(\frac{\lambda'}{2R^2}-\frac{\lambda}{R^3}R').
\end{eqnarray*}

The non-zero components of $F_{ab}$ are:
\begin{eqnarray*}
F_{01}=\frac{1}{R^4}(Q(r^2-N^2)+2HNr) = -\frac{Q}{R^2} +2r \frac{Q r +NH}{R^4}
\nonumber \\
F_{23}=\frac{1}{R^4}(H(r^2-N^2)-2QNr) = \frac{H}{R^2} - 2N \frac{Q r +NH}{R^4}
\end{eqnarray*}
so that
\begin{eqnarray}
F_{ab} \gamma^{ab} & = & -2 F_{01} \gamma_{01} +2 F_{23} \gamma_{23}
\nonumber\\
&= & -\frac{2}{R^4} \gamma_{01} (r+\gamma_5 N)^2 (Q - \gamma_5 H).
\end{eqnarray}

We now compute the expressions for $\omega_\mu^{ab}\gamma_{ab}$:
\beqs
\omega_t^{ab}\gamma_{ab} &=& \frac{2}{R^4} \gamma_{01}
\left[ (r-M)R^2 - \lambda (r+\gamma_5 N) \right], \\
\omega_r^{ab}\gamma_{ab} &=& 0 , \\
\omega_\theta^{ab}\gamma_{ab} &=& -2 \frac{\sqrt{\lambda}}{R^2}
\gamma_{12} (r+\gamma_5 N), \\
\omega_\phi^{ab}\gamma_{ab} &=& -2 \frac{\sqrt{\lambda}}{R^2}
\sin\theta \gamma_{13} (r+\gamma_5 N) -2\cos\theta \gamma_{23}\nonumber \\
& &
+4N \cos\theta \frac{1}{R^4} \gamma_{01}\left[ (r-M)R^2 - \lambda (r
+\gamma_5 N)\right].
\eeqs
Taking also into account that
\beq
\gamma_t=\frac{\sqrt{\lambda}}{R}\gamma_0, \quad
\gamma_r= \frac{R}{\sqrt{\lambda}}\gamma_1, \quad
\gamma_\theta= R\gamma_2, \quad
\gamma_\phi= R\sin\theta \gamma_3 +2N \frac{\sqrt{\lambda}}{R}\cos\theta
\gamma_0,
\eeq
we finally arrive at the SUSY variations
\beqs
\delta \psi_t & = & \partial_t \epsilon + \frac{1}{2R^4} \gamma_{01}
\left\{(r-M)R^2 - \lambda (r+\gamma_5 N) - i(r+\gamma_5 N)^2 (Q-\gamma_5 H)
\frac{\sqrt{\lambda}}{R}\gamma_0\right\}\epsilon ,\nonumber \\
\delta \psi_r & = & \partial_r \epsilon - i \frac{1}{2R^4} \gamma_{01}
(r+\gamma_5 N)^2 (Q-\gamma_5 H) \frac{R}{\sqrt{\lambda}}\gamma_1 \epsilon,
\nonumber\\
\delta \psi_\theta & = & \partial_\theta \epsilon + \frac{1}{2R^4}
\left\{-\sqrt{\lambda} R^2 \gamma_{12}(r+\gamma_5 N) \nonumber
 - i\gamma_{01}(r+\gamma_5 N)^2 (Q-\gamma_5 H) R\gamma_2\right\}\epsilon ,\\
\delta \psi_\phi & = & \partial_\phi \epsilon + \frac{1}{2R^4}
\left\{-\sqrt{\lambda} R^2 \sin\theta \gamma_{13} (r+\gamma_5 N)
-R^4 \cos\theta \gamma_{23} \right. \nonumber\\ & & \qquad
+2N \cos\theta \gamma_{01}\left[ (r-M)R^2 - \lambda (r+\gamma_5 N)\right]
\nonumber \\ & & \qquad \left.
- i\gamma_{01}(r+\gamma_5 N)^2 (Q-\gamma_5 H)\left(R\sin\theta \gamma_3
+2N \frac{\sqrt{\lambda}}{R}\cos\theta \gamma_0\right)\right\}\epsilon.
\eeqs

Note that in flat space we still have non trivial equations:
\begin{eqnarray}\label{kilspineq3}
&& \partial_t \epsilon = 0 \nonumber \\
&& \partial_r \epsilon = 0 \nonumber\\
&& \partial_{\theta} \epsilon = \frac{1}{2}\gamma_{12}\epsilon \nonumber\\
&& \partial_{\phi} \epsilon = \frac{1}{2} (\sin\theta \:\gamma_{13}
+  \cos\theta \:\gamma_{23})\epsilon
\end{eqnarray}
The general expression for the Killing spinor satisfying equations
(\ref{kilspineq3})
is:
\begin{eqnarray}
\epsilon(t,r,\theta,\phi)=
e^{\frac{1}{2}\gamma_{12}\theta}\:
e^{\frac{1}{2}\gamma_{23}\phi} \epsilon_0 \label{sphkill2}
\end{eqnarray}
where $\epsilon_0$ is a constant spinor.

In our more general case, let us suppose that all the dependence
in $\theta$ and $\phi$ factorizes as above. Hence, we look for a Killing
spinor with the form (\ref{sphkill2}) where however $\epsilon_0$ depends
on $r$ and possibly $t$.

Let us first look at the expression for $\delta \psi_\theta$. It becomes
an algebraic condition on $\epsilon_0$, which can be rewritten as:
\beq
\left[(r+\gamma_5 N)(\sqrt{\lambda} -r- \gamma_5 N)\gamma_0
-i (Q+\gamma_5 H) R \right] \epsilon_0 \equiv P \epsilon_0 = 0,
\eeq
where we have used $R^2=r^2+N^2=(r+\gamma_5 N)(r-\gamma_5 N)$.

The Killing spinor equations will have non-trivial
solutions only if the operator $P$ above has vanishing eigenvalues, i.e.
its determinant is zero.
It appears however easier to just compute the square of the operator $P$:
\beq
P^2= (-2iQ R)  P -R^2 \left[ (\sqrt{\lambda} -r)^2 +N^2 -Q^2 -H^2\right].
\eeq
The coefficients are just complex numbers, so that the eigenvalues of $P$
must satisfy the same equation, with two solutions.
Therefore, the operator $P$ will have zero eigenvalues (and be proportional
to a projector) only
if $(\sqrt{\lambda} -r)^2 +N^2 -Q^2 -H^2 = 0$, which translates into
\beq
M^2+N^2=Q^2+H^2, \label{bpsbound2}
\eeq
an $r$-independent condition.
Note that another way to state the above BPS condition is to write
$\sqrt{\lambda} = r-M$. It is this expression that
we will substitute back into the SUSY variations. This is done in
Section 2.

\section{Computation of the Killing spinor}
In this section we compute the explicit expression of the Killing spinor,
using the results obtained in Section 2, namely that the Killing spinor
has to satisfy the projection
\beq
\left\{1-ie^{(\beta+\alpha_m-\alpha_q)\gamma_5}\gamma_0\right\}\epsilon=0.
\label{proj2}
\eeq

The only non trivial equation that remains to be solved is $\delta\psi_r=0$:
\beq
\partial_r \epsilon = \frac{Z}{2R(r-M)} i
e^{(2\beta-\alpha_q)\gamma_5}\gamma_0\epsilon.
\eeq
The strategy we adopt is straightforward. We just solve the
projector equation above in components, and then plug back the
components into the first order differential equation.

Let us call
\beq
c\equiv \cos(\beta+\alpha_m-\alpha_q), \qquad s \equiv \sin
(\beta+\alpha_m-\alpha_q).
\eeq
Then the solution to (\ref{proj2}) can be written in the form
(\ref{sphkill2}) with
\beq
\epsilon_0 = \epsilon_1(r) \left(\begin{array}{c}1\\ 0 \\ is \\ ic
\end{array} \right)
+\epsilon_2(r) \left(\begin{array}{c}0 \\ 1 \\ -ic \\ is \end{array}\right).
\eeq
The equation $\delta\psi_r=0$ becomes then
\beq
\partial_r \epsilon_0 = \frac{Z}{2R(r-M)}\left(\cos(\beta-\alpha_m)
+\sin(\beta-\alpha_m) \gamma_5\right) \epsilon_0.
\eeq
In computing $\partial_r \epsilon_0$, one has to recall that
$\partial_r \beta = -N/R^2$. We have 4 equations for 2 functions
$\epsilon_1(r)$ and $\epsilon_2(r)$. It is fairly straightforward to
see that the equations for the two lower components of $\epsilon_0$
are automatically satisfied once the equations for the upper two
components are satisfied.

The two equations to be solved are
\beqs
\partial_r \epsilon_1 &= &  \frac{Z}{2R(r-M)}\left(\cos(\beta-\alpha_m)
\epsilon_1 - \sin(\beta-\alpha_m) \epsilon_2\right), \\
\partial_r \epsilon_2 &= &  \frac{Z}{2R(r-M)}\left(\sin(\beta-\alpha_m)
\epsilon_1 + \cos(\beta-\alpha_m) \epsilon_2\right).
\eeqs
We can clearly write the 2 functions $\epsilon_1(r)$ and $\epsilon_2(r)$
in terms of a common scalar function and a phase:
\beq
\left(\begin{array}{c} \epsilon_1(r) \\ \epsilon_2(r) \end{array} \right)
=h(r) \left(\begin{array}{cc} \cos \hat \alpha(r) & \sin \hat \alpha(r) \\
-\sin \hat \alpha(r) & \cos \hat \alpha(r) \end{array} \right)
\left(\begin{array}{c} \hat \epsilon_1 \\ \hat \epsilon_2 \end{array} \right)
\eeq
where $\hat \epsilon_1$ and $\hat \epsilon_2$ are constants.

We obtain the two equations
\beq
\partial_r h= \frac{Z}{2R(r-M)} \cos(\beta-\alpha_m) h , \qquad
\partial_r \hat \alpha = - \frac{Z}{2R(r-M)} \sin(\beta-\alpha_m).
\eeq
They can be rewritten as
\beq
\partial_r h= \frac{N^2+rM}{2R^2(r-M)} h , \qquad
\partial_r \hat \alpha = \frac{N}{2R^2} \equiv -\frac{1}{2}\partial_r \beta.
\eeq
The solution is thus:
\beq
\left(\begin{array}{c} \epsilon_1(r) \\ \epsilon_2(r) \end{array} \right)
=\left(\frac{r-M}{R}\right)^{\frac{1}{2}}
\left(\begin{array}{cc} \cos \frac{1}{2} \beta(r) & -\sin \frac{1}{2}\beta(r)
\\
\sin \frac{1}{2} \beta(r) & \cos\frac{1}{2} \beta(r)  \end{array} \right)
\left(\begin{array}{c} \hat \epsilon_1 \\ \hat \epsilon_2 \end{array} \right).
\eeq
We can define the rotation matrix
\beq
R[\alpha]= \left(\begin{array}{cc} \cos \alpha & -\sin \alpha \\
\sin  \alpha & \cos  \alpha \end{array} \right).
\eeq
Then by performing an additional constant rotation of the spinors
\beq
\left(\begin{array}{c} \hat \epsilon_1 \\ \hat \epsilon_2 \end{array} \right)
\equiv \vec{\hat \epsilon} = \begin{array}{c} R[\frac{1}{2}(\alpha_m-\alpha_q)]
\end{array}
\vec{\epsilon}
\eeq
we can write the final expression for the Killing spinor as:
\beq
\epsilon_0(r) =  \left(\frac{r-M}{R}\right)^{\frac{1}{2}}
\left(\begin{array}{c} R[\frac{1}{2}(\beta(r)+\alpha_m-\alpha_q)]
\vec{\epsilon} \\
i R[\frac{1}{2}(\pi -\beta(r)-\alpha_m+\alpha_q)]
\vec{\epsilon} \end{array} \right).
\eeq
This is the expression presented in Section 2.

\end{document}